# Co-Optimization Generation and Distribution Planning in Microgrids

Hossein Lotfi, Amin Khodaei
Department of Electrical and Computer Engineering
University of Denver
Denver, CO, USA
hossein.lotfi@du.edu, amin.khodaei@du.edu

*Abstract*—This paper proposes a co-optimization generation and distribution planning model in microgrids in which simultaneous investment in generation, i.e., distributed generation (DG) and distributed energy storage (DES), and distribution, i.e., upgrading the existing distribution network, is considered. The objective of the proposed model is to minimize the microgrid total planning cost which comprises the investment cost of installed generation assets and lines, the microgrid operation cost, and the cost of unserved energy. The microgrid planning solution determines the optimal generation size, location, and mix, as well as required network upgrade. To consider line flow and voltage limits, a linearized power flow model is proposed and used, allowing further application of mixed integer linear programming (MILP) in problem modeling. The proposed model is applied to the IEEE 33-bus standard test system to demonstrate the acceptable performance and the effectiveness of the proposed model.

## NOMENCLATURE

*Indices:*
| | |
|---|---|
| $ch$ | Superscript for DES charging mode |
| $d$ | Index for day |
| $dch$ | Superscript for DES discharging mode |
| $h$ | Index for hour |
| $i$ | Index for DERs |
| $l$ | Index for lines |
| $m, n$ | Index for buses |
| $s$ | Index for scenarios |
| $t$ | Index for year |

*Sets:*
| | |
|---|---|
| $B$ | Set of buses |
| $B_l$ | Set of buses at both ends of line $l$ |
| $E$ | Set of DES units |
| $G$ | Set of dispatchable DGs |
| $L$ | Set of lines |
| $L_m$ | Set of lines connected to bus $m$ |
| $W$ | Set of nondispatchable DGs |

*Parameters:*
| | |
|---|---|
| $a_1/a_2$ | Line-bus connection indicator (1 if connected at from/to bus, 0 otherwise) |
| $b$ | Line susceptance |
| $c$ | Generation price for dispatchable DGs |
| $C^{cap}$ | Allowable DES installation capacity |
| $CC$ | Annualized DG investment cost |
| $CE$ | Annualized DES investment cost – energy |
| $CL$ | Annualized line investment cost |
| $CP$ | Annualized DES investment cost – power |
| $g$ | Line conductance |
| $M$ | Large positive constant |
| $P^{cap}$ | Allowable DER installation capacity |
| $P^{M,max}$ | Flow limit between microgrid and the utility grid |
| $PD$ | Active load |
| $PD^{max}$ | Maximum demand during the planning horizon |
| $PL^{max}$ | Active line flow limit |
| $pr$ | Probability of each scenario |
| $QL^{max}$ | Reactive line flow limit |
| $QD$ | Reactive load |
| $r$ | Discount rate |
| $u$ | Binary islanding parameter (1 if grid connected, 0 if islanded) |
| $v$ | Value of lost load (VOLL) |
| $\beta$ | Ratio of critical loads to total load |
| $\eta$ | DES efficiency |
| $\kappa$ | Coefficient of present-worth value |
| $\mu$ | Normalized generation forecast of nondispatchable DGs |
| $\rho$ | Market price |

*Variables:*
| | |
|---|---|
| $C^{max}$ | Installed DES capacity |
| $IC$ | Microgrid investment cost |
| $LS$ | Load shedding |
| $OC$ | Microgrid operation cost |
| $P$ | DER active power output |
| $P^{ch}$ | DES charging power |
| $P^{dch}$ | DES discharging power |
| $P^M$ | Microgrid exchanged power with the utility grid |
| $P^{max}$ | Installed DER capacity |
| $PL$ | Active line flow |
| $Q$ | DER reactive power output |
| $QL$ | Reactive line flow |
| $RC$ | Microgrid reliability cost |
| $V$ | Bus voltage magnitude |
| $x$ | DER investment state (1 if installed, 0 otherwise) |
| $z$ | Line investment state (1 if installed, 0 otherwise) |
| $\theta$ | Bus voltage angle |

## I. INTRODSUCTION

Microgrids, as modern small-scale power systems capable of islanding and self-supply, offer many advantages for grid operators and consumers, such as improving reliability by islanding and decreasing outage time, improving power quality by managing local loads, providing better operational economics by selling energy back to the utility grid and reducing power delivery costs, and reducing carbon emission by utilizing renewable distributed generations (DGs) [1-5].

Congestion in distribution lines or disturbance in the upstream utility grid may prevent fully supplying the loads in a distribution network. Moreover, addition of new loads to the network may require timely upgrade of the existing distribution network assets. An efficient distribution planning is required in this case to cope with the potential network problems. There are various methodologies proposed in the literature for distribution network planning. In [6], a methodology for optimal expansion planning of distribution networks is presented which considers network contingencies and relocation of switchgears. The optimization methodology consists of two stages in which the investment and operation problems are solved in the first and second stages, respectively. The study in [7] proposes an algorithm to capture the load variations along with the generation volatility and intermittency of renewable energy sources. The proposed model coordinates voltage control among smart grid technologies by determining the optimal number of DG units. In [8], a model for distribution grid planning enhancement is presented using profiling estimation technique. The objective of the proposed model is to reconstruct the load profile of the medium/low voltage substations. The study in [9] proposes a methodology to be used by distribution system operators (DSOs) for optimal distribution grid planning. The proposed model can be used in meshed and radial grids. Both passive and active network measures are considered in this study. The solution of this model determines whether a new line or transformer should be installed or any other reinforcement actions should be taken. In [10], the economic impact of demand response on distribution network planning is investigated. The reference network model, a large-scale distribution network planning tool, is used to take appropriate action in response to demand growth in a ten-year planning horizon. The study in [11] presents the microgrid planning as an alternative to generation and transmission expansion. The microgrid-based co-optimization planning problem is solved by decomposition to a planning problem and annual reliability subproblem. Microgrid planning is also investigated in [12]-[15]. The study in [12] proposes microgrid planning under uncertainty, and solves the planning problem by decomposition to an investment master problem and an operation subproblem. In [13]-[15], individual AC/DC and hybrid AC/DC microgrid planning problems are discussed in which the optimal distributed energy resource (DER) size and location as well as the type of the microgrid are determined, but power flow and line losses are overlooked.

This paper presents a co-optimization generation and distribution planning in a microgrid aiming at minimizing the microgrid long-term operation cost while ensuring a reliable supply of loads. One solution to increase the distribution network reliability and prevent load curtailment is to build new distribution lines or to reinforce the existing lines through upgrades. Another solution is to install DERs in strategic locations in distribution network. In this paper, both these solutions are considered simultaneously, allowing the identification of the most viable solution. Various types of DERs are considered in this study in which their optimal size and location are determined through the proposed model. The power flow equations are linearized, using minor approximations, in order to be able to formulate the problem using mixed integer linear programming (MILP). The rest of the paper is organized as follows: Section II presents the outline and formulates the microgrid co-optimization generation and distribution planning problem. Section III presents numerical simulations on a test microgrid, and Section IV concludes the paper.

## II. MODEL OUTLINE AND FORMULATION

There are both dispatchable and nondispatchable candidate DGs in the microgrid. Nondispatchable DGs are renewable energy sources such as solar PV and wind. Distributed energy storage (DES) is employed in order to increase the controllability and dispatchability of these energy sources. DES is charged at off-peak hours with low electricity prices and discharged at peak hours when electricity price is high. The microgrid is connected to the utility grid to exchange power as needed and further govern voltage and frequency. One significant feature of the microgrid is its islanding capability which allows operation in the islanded mode in case of any disturbance in the upstream grid. Islanding is defined as a set of scenarios in the planning problem as will be further explained. The microgrid can buy power from the utility grid, associated with positive exchanged power, or sell back the excess power to the utility grid, associated with negative exchanged power which increases the microgrid revenue. A number of candidate distribution lines between predetermined buses are considered in order to alleviate potential congestion in existing lines. The solution of the optimization problem determines the optimal size and location of DERs as well as the installation of lines.

The proposed co-optimization generation and distribution planning problem aims at minimizing the microgrid total planning cost (1) comprising of the investment cost of DERs and distribution lines (IC), the operation cost (OC), and the reliability cost (RC). A discount rate $r$ is considered in order to evaluate the objective in terms of discounted costs, which is appeared in the objective as the present-worth value $\kappa_t$, where $\kappa_t = 1/(1+r)^{t-1}$. It should be noted that the investment cost is calculated annually while the operation and reliability costs are calculated hourly for all hours and days in the planning horizon. The investment, operation, and reliability costs are defined in (2)-(4), respectively.

$$\min \sum_t \kappa_t \left( IC_t + OC_t + RC_t \right) \quad (1)$$

$$IC_t = \sum_{i \in \{G,W\}} CC_{it} P_i^{\max} + \sum_{i \in E} \left( CP_{it} P_i^{\max} + CE_{it} C_i^{\max} \right) + \sum_{l \in L} CL_{lt} z_l \quad \forall t \quad (2)$$

$$OC_t = \sum_d \sum_h \sum_{i \in G} c_i P_{ihdt0} + \sum_d \sum_h \rho_{hdt} P_{hdt0}^M \quad \forall t \quad (3)$$

$$RC_t = \sum_s pr_s \sum_d \sum_h \sum_m vLS_{mhdts} \quad \forall t \quad (4)$$

The investment cost (2) comprises the investment cost of dispatchable and nondispatchable DGs (derived by multiplying the DGs' annualized capital cost by their installed capacity), investment cost of the DES, and investment cost of distribution lines. The DES investment cost has two components associated with installed power capacity and energy capacity, in which both are calculated as the associated annualized capital cost times installed capacity. The investment cost of line is determined as the given annualized capital cost times a binary investment variable, $z_l$. The binary variable is employed to consider the installation of distribution lines; that is if a candidate line is installed, $z_l$ would be one, otherwise it is zero. The operation cost (3) consists of two terms, the operation cost of dispatchable DGs calculated by their generation price times generated power in each hour, and cost of power exchanged with the utility grid, calculated by electricity market price times the amount of exchanged power with the utility grid. Both terms are aggregated over all hours and days in the planning horizon. The reliability cost (4) represents the cost of unserved energy and is defined as the value of lost load (VOLL) times the amount of hourly load curtailment, aggregated over all hours, days, and islanding scenarios in the planning horizon. A comprehensive discussion on VOLL for different types of customers can be found in [13]. The operation and reliability costs are further summed over the considered scenarios (for grid-connected and islanded operation) based on the associated probability. In (3), $s=0$ represents the grid-connected mode. The objective function (1) is further subject to DERs and power balance constraints (5)-(19) and power flow equations (20)-(28).

**DERs and Power Balance Constraints:** A binary decision variable, $x_{im}$, is used to determine the location of DER installations, which would be one when DER $i$ is installed at bus $m$, and zero otherwise. Constraint (5) ensures that each DER is connected to only one bus. The total dispatchable capacity should be larger than the microgrid critical load to ensure a reliable supply of loads when operating in the islanded mode (6). The active load balance equation (7) ensures that the generated power from all DERs and lines connected to each bus plus the exchanged power with the utility grid at the point of interconnection (POI) is equal to the hourly load demand minus the amount of curtailed load. Similarly, the reactive load balance equation (8) ensures that the reactive power from all DERs and lines connected to each bus plus the exchanged reactive power with the utility grid is equal to the amount of hourly reactive load. The exchanged power with the utility grid is limited by the capacity of the line connecting the microgrid to the utility grid (9). The amount of hourly generated power of dispatchable DGs cannot exceed their installed capacity (10). The hourly power generated by nondispatchable DGs is determined by a normalized forecasted generation times the associated installed capacity (11). Additionally, installed DG capacity cannot exceed its allowable installation capacity limits (12), which is determined based on budget or space limitations. The load shedding at each bus cannot exceed its hourly load demand (13). The DES constraints are represented in (14)-(19). The DES power in both discharging and charging modes is limited by its installed power capacity (14)-(15). The DES stored energy is determined based on the net charged power, efficiency, and the stored energy in previous hours (16). Additionally, the DES net charge is assumed to be zero at the end of each day in the planning horizon (17). Finally, the installed DES power and energy capacity are limited by its allowable power and energy capacity limits, respectively (18)-(19).

$$\sum_m x_{im} \leq 1 \quad \forall i \in \{G, W, E\} \quad (5)$$

$$\beta \sum_m PD_{mhdt}^{\max} \leq \sum_{i \in G} P_i^{\max} \quad (6)$$

$$\sum_{i \in \{G,W\}} P_{ihdts} x_{im} + \sum_{i \in E} \left( P_{ihdts}^{dch} - P_{ihdts}^{ch} \right) x_{im} + \sum_{l \in L_m} PL_{lhdts} + P_{mhdts}^M$$
$$= PD_{mhdt} - LS_{mhdt} \quad \forall m, \forall h, \forall d, \forall t, \forall s \quad (7)$$

$$\sum_{i \in \{G,W\}} Q_{ihdts} x_{im} + \sum_{l \in L_m} QL_{lhdts} + Q_{mhdts}^M = QD_{mhdt}$$
$$\forall m, \forall h, \forall d, \forall t, \forall s \quad (8)$$

$$-P^{M,\max} u_{hdts} \leq P_{hdts}^M \leq P^{M,\max} u_{hdts} \quad \forall h, \forall d, \forall t, \forall s \quad (9)$$

$$0 \leq P_{ihdts} \leq P_i^{\max} \quad \forall i \in G, \forall h, \forall d, \forall t, \forall s \quad (10)$$

$$P_{ihdts} = P_i^{\max} \mu_{ihdt} \quad \forall i \in W, \forall h, \forall d, \forall t, \forall s \quad (11)$$

$$P_i^{\max} \leq P_i^{cap} \sum_m x_{im} \quad \forall i \in \{G, W\} \quad (12)$$

$$LS_{mhdts} \leq PD_{mhdts} \quad \forall m, \forall h, \forall d, \forall t, \forall s \quad (13)$$

$$0 \leq P_{ihdts}^{dch} \leq P_i^{\max} \quad \forall i \in E, \forall h, \forall d, \forall t, \forall s \quad (14)$$

$$0 \leq P_{ihdts}^{ch} \leq P_i^{\max} \quad \forall i \in E, \forall h, \forall d, \forall t, \forall s \quad (15)$$

$$0 \leq \sum_{k \leq h} (P_{ikdts}^{ch} - P_{ikdts}^{dch}/\eta_i) \leq C_i^{\max} \quad \forall i \in E, \forall h, \forall d, \forall t, \forall s \quad (16)$$

$$\sum_h (P_{ihdts}^{ch} - P_{ihdts}^{dch}/\eta_i) = 0 \quad \forall i \in E, \forall d, \forall t, \forall s \quad (17)$$

$$P_i^{\max} \leq P_i^{cap} \sum_m x_{im} \quad \forall i \in E \quad (18)$$

$$C_i^{\max} \leq C_i^{cap} \sum_m x_{im} \quad \forall i \in E \quad (19)$$

**Power Flow Constraints:** Power flow equations are nonlinear and cannot be directly included in the developed MILP formulation. In order to linearize the equations, (20) and (21) are assumed. Voltage magnitudes and angles are considered as those of bus 1 (i.e., the POI) plus deviations, as represented in (22) and (23). The resulting multiplication of voltage magnitude and voltage angle variables is very small and thus can be eliminated from power flow equations.

$$\sin(\theta_{mhdts} - \theta_{nhdts}) \approx \theta_{mhdts} - \theta_{nhdts} \quad \forall mn \in L, \forall h, \forall d, \forall t, \forall s \quad (20)$$

$$\cos(\theta_{mhdts} - \theta_{nhdts}) \approx 1 \quad \forall mn \in L, \forall h, \forall d, \forall t, \forall s \quad (21)$$

$$V_{mhdts} = 1.0 + \Delta V_{mhdts} \quad \forall m, \forall h, \forall d, \forall t, \forall s \quad (22)$$

$$\theta_{mhdts} = 0 + \Delta \theta_{mhdts} \quad \forall m, \forall h, \forall d, \forall t, \forall s \quad (23)$$

The linear active and reactive power flow equations for distribution lines are represented by (24) and (25), respectively. If a candidate line is not installed, $z_l$ would be zero, and (24)-(25) would be relaxed. Therefore, the real and reactive powers passing through the lines would be zero according to (26) and (27). Likewise, if the solution of the optimization problem is to install a line, $z_l$ would be one, and real and reactive powers would be respectively determined by (24) and (25) with the limits imposed by (26) and (27). It should be noted that (24) and (25) are not linear, and are solved in a two-stage fashion.

The term $\sum_{m\in B_l} a_{lm} \Delta V_{mhdts}$ is considered zero in stage one, and then by finding $\Delta V_{mhdts}$, this term will be replaced, and the problem is solved again in stage two. Finally, the voltage magnitudes at all buses cannot exceed their minimum and maximum limits (28).

$$-M(1-z_l) \leq PL_{lhdts} - g_l(1+\sum_{m\in B_l} a_{lm1}\Delta V_{mhdts})\sum_{m\in B_l}((a_{lm2}-a_{lm1})\Delta V_{mhdts}) +$$
$$b_l \sum_{m\in B_l}(a_{lm2}-a_{lm1})\Delta\theta_{mhdts} \leq M(1-z_l) \qquad \forall l, \forall h, \forall d, \forall t, \forall s \quad (24)$$

$$-M(1-z_l) \leq QL_{lhdts} + b_l(1+\sum_{m\in B_l} a_{lm1}\Delta V_{mhdts})\sum_{m\in B_l}((a_{lm2}-a_{lm1})\Delta V_{mhdts}) +$$
$$g_l \sum_{m\in B_l}(a_{lm2}-a_{lm1})\Delta\theta_{mhdts} \leq M(1-z_l) \qquad \forall l, \forall h, \forall d, \forall t, \forall s \quad (25)$$

$$-PL_l^{max} z_l \leq PL_{lhdts} \leq PL_l^{max} z_l \qquad \forall l, \forall h, \forall d, \forall t, \forall s \quad (26)$$

$$-QL_l^{max} z_l \leq QL_{lhdts} \leq QL_l^{max} z_l \qquad \forall l, \forall h, \forall d, \forall t, \forall s \quad (27)$$

$$\Delta V_m^{min} \leq \Delta V_{mhdts} \leq \Delta V_m^{max} \qquad \forall m, \forall h, \forall d, \forall t, \forall s \quad (28)$$

## III. NUMERICAL SIMULATIONS

The IEEE standard 33-bus test system, as shown in Fig. 1 is used for microgrid installation. This system comprises 33 buses, 32 distribution lines, and 32 loads, with a maximum initial aggregated load of 2.7 MW [14]. Tables I, II and III show the characteristics of candidate DGs, DES, and distribution lines, respectively. As renewable DGs have a negligible operation cost, their cost coefficient is assumed to be zero. The investment cost of the candidate lines is calculated based on studies in [16]. The hourly load demand, renewable generation, and market price data are forecasted based on the historical data from a practical system [14]. The DES efficiency is assumed to be 95%. The planning horizon is 20 years. No islanding scenarios are considered in simulations, meaning that the microgrid operates in grid-connected mode at all times. However, the proposed model can efficiently consider islanded operation. The microgrid planning problem is implemented on a high performance computing server consisting of four 10-core Intel Xeon E7-4870 2.4 GHz processors. The problem is formulated by MILP and solvedtabl by CPLEX 12.6 [17], with average running time of 70 minutes. Following cases are studied.

Case 0: Base case microgrid planning
Case 1: Sensitivity analysis on the ratio of critical loads
Case 2: Sensitivity analysis on load demand
Case 3: Sensitivity analysis on market prices

**Case 0:** The ratio of critical loads to total load is considered to be 40% for all operation hours. It is assumed that DGs 1-6 can be installed in buses 17, 21, 32, 24, 15, and 15, respectively, as end lines have lower capacity and congestion is more likely. It is further assumed that the DES can be installed in bus 15. The microgrid planning solution would install dispatchable DGs 3 (with 0.65 MW capacity) and 4 (with 0.44 MW capacity) as well as the solar unit (with 0.48 MW capacity). No candidate lines are installed in the base case. The total planning cost is $9,462,578 with a cost breakdown of $1,310,805 for the investment cost and $8,151,773 for the operation cost.

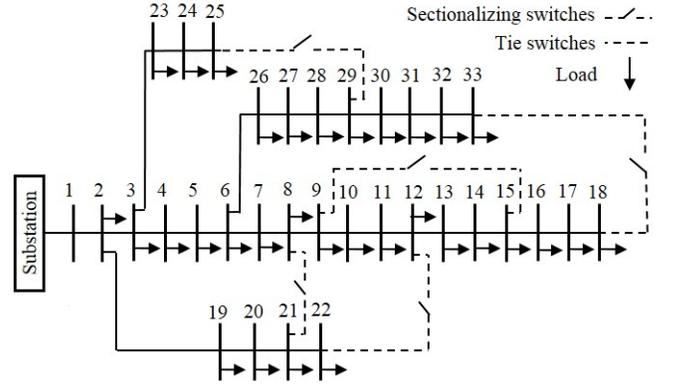

Fig. 1. IEEE 33-bus test system.

TABLE I
CANDIDATE DGs CHARACTERISTICS

| Unit No. | Type | Allowable installation capacity (MW) | Cost Coefficient ($/MWh) | Annualized Investment Cost ($/MW) |
|---|---|---|---|---|
| 1 | Gas | 3 | 90 | 50,000 |
| 2 | Gas | 3 | 90 | 50,000 |
| 3 | Gas | 1 | 70 | 70,000 |
| 4 | Gas | 1 | 70 | 70,000 |
| 5 | Wind | 2 | 0 | 132,000 |
| 6 | Solar | 2 | 0 | 133,000 |

TABLE II
CANDIDATE DES CHARACTERISTICS

| Allowable Installation Capacity (MW) | Allowable Installation Energy (MWh) | Annualized Investment Cost – Power ($/MW) | Annualized Investment Cost – Energy ($/MWh) |
|---|---|---|---|
| 1 | 6 | 60,000 | 30,000 |

TABLE III
CANDIDATE LINES CHARACTERISTICS

| Line | From bus | To bus | R(Ω) | X(Ω) | Line Capacity (kW) | Annualized Investment Cost ($) |
|---|---|---|---|---|---|---|
| 33 | 12 | 13 | 1.468 | 1.155 | 500 | 37749 |
| 34 | 13 | 14 | 0.5416 | 0.7129 | 450 | 12534 |
| 35 | 14 | 15 | 0.591 | 0.526 | 300 | 9118 |
| 36 | 15 | 16 | 0.7463 | 0.545 | 250 | 9595 |
| 37 | 16 | 17 | 1.289 | 1.721 | 250 | 16573 |
| 38 | 17 | 18 | 0.732 | 0.574 | 100 | 3765 |
| 39 | 20 | 21 | 0.4095 | 0.4784 | 210 | 4423 |
| 40 | 21 | 22 | 0.7089 | 0.9373 | 110 | 4010 |
| 41 | 23 | 24 | 0.898 | 0.7091 | 1050 | 48492 |
| 42 | 24 | 25 | 0.896 | 0.7011 | 500 | 23040 |
| 43 | 30 | 31 | 0.9744 | 0.963 | 500 | 25056 |

**Case 1:** The impact of changing the ratio of critical loads, $\beta$, on planning results is studied in this case. The microgrid planning is studied two scenarios, with and without allowing installation of candidate lines, and results are tabulated in Table IV. The total dispatchable capacity increases by increasing the ratio of critical loads as the microgrid should be able to seamlessly supply critical loads. Therefore, the microgrid investment cost increases too by increasing $\beta$, as shown in Table V. By increasing the ratio of critical loads from 0 to 60%, none of the candidate lines are installed, but by increasing $\beta$ to 0.8 and 1 (meaning all loads are considered as critical), lines 34, 35, and 39 are installed. The investment cost suddenly increases in comparison with the case without line installation. If there is no critical load in the microgrid (associated with $\beta=0$), only the solar unit is installed, but none of the dispatchable units, meaning that importing power from the utility grid is more economical than installing local DGs. It is worth mentioning that for the ratio of critical loads at 80% and 100%, a larger capacity of solar unit is also installed when the line installation is considered (increased from 0.48 MW to 0.78MW). The reason is that the solar unit is installed in bus 15, and lines 34 and 35 are respectively between buses 13-14 and 14-15, thus can help with transferring the additional generated power. As total load does not change, there would be excess power to sell back to the utility grid, which causes the operation cost to decrease. As the increase in the investment cost is higher than the decrease in the operation cost, the planning cost would increase by increasing the ratio of critical loads. According to results, the planning cost would decrease in case of the installation of candidate lines, which means the simultaneous installation of DERs and distribution lines would be more economical. It should be noted that DES is not installed for any ratio of critical loads.

TABLE IV
INVESTMENT PLAN WITH RESPECT TO CHANGES IN RATIO OF CRITICAL LOADS

| $\beta$ | | 1 | 2 | 3 | 4 | 5 | 6 | Installed Lines |
|---|---|---|---|---|---|---|---|---|
| 0 | w/o lines | 0 | 0 | 0 | 0 | 0 | 0.48 | - |
|   | w/ lines  | 0 | 0 | 0 | 0 | 0 | 0.48 | - |
| 0.2 | w/o lines | 0 | 0 | 0 | 0.55 | 0 | 0.48 | - |
|     | w/ lines  | 0 | 0 | 0 | 0.55 | 0 | 0.48 | - |
| 0.4 | w/o lines | 0 | 0 | 0.10 | 1.00 | 0 | 0.48 | - |
|     | w/ lines  | 0 | 0 | 0.10 | 1.00 | 0 | 0.48 | - |
| 0.6 | w/o lines | 0 | 0 | 0.65 | 1.00 | 0 | 0.48 | - |
|     | w/ lines  | 0 | 0 | 0.65 | 1.00 | 0 | 0.48 | - |
| 0.8 | w/o lines | 0.17 | 0.35 | 0.68 | 1.00 | 0 | 0.48 | - |
|     | w/ lines  | 0.17 | 0.35 | 0.66 | 1.00 | 0 | 0.78 | 34,35,39 |
| 1.0 | w/o lines | 0.28 | 0.76 | 0.68 | 1.00 | 0 | 0.48 | - |
|     | w/ lines  | 0.50 | 0.55 | 0.68 | 1.00 | 0 | 0.78 | 34,35,39 |

TABLE V
MICROGRID COSTS WITH RESPECT TO RATIO OF CRITICAL LOADS

| $\beta$ | | Investment Cost ($) | Operation Cost ($) | Planning Cost ($) |
|---|---|---|---|---|
| 0 | w/o lines | 596650 | 8809487 | 9406137 |
|   | w/ lines  |        |         |         |
| 0.2 | w/o lines | 9406137 | 8459261 | 9412997 |
|     | w/ lines  |         |         |         |
| 0.4 | w/o lines | 1310805 | 8151773 | 9462578 |
|     | w/ lines  |         |         |         |
| 0.6 | w/o lines | 1668653 | 7864340 | 9532992 |
|     | w/ lines  |         |         |         |
| 0.8 | w/o lines | 1932385 | 7759704 | 9692089 |
|     | w/ lines  | 2354132 | 7116085 | 9470217 |
| 1.0 | w/o lines | 2187500 | 7758661 | 9946161 |
|     | w/ lines  | 2613157 | 7082358 | 9695515 |

**Case 2:** In this case, a sensitivity analysis of planning results with respect to the load demand is carried out. The installed DER capacity and installed lines are represented in Table VI. The hourly load in all years is increased by up to 100%, investigating additional cases with different load growth rates. As expected, by increasing the load demand, more DER capacity should be installed, which causes the investment cost to increase (Table VII). It should be noted that among dispatchable DGs, units 3 and 4, despite their higher capital costs, are installed first because they are associated with a lower cost coefficient compared to that of units 1 and 2. Also, following more than 60% increase in the load, the microgrid planning solution would install candidate lines, as represented in Table VI, which causes a sudden increase in the investment cost. On the other hand, by increasing the total load, more power is generated by DGs, and more power would be imported from the utility grid, which cause the operation cost, and hence the planning cost, to increase.

TABLE VI
INVESTMENT PLAN WITH RESPECT TO LOAD CHANGES

| Load change percentage | 1 | 2 | 3 | 4 | 5 | 6 | Installed Lines |
|---|---|---|---|---|---|---|---|
| Orig. | 0 | 0 | 0.10 | 1.00 | 0 | 0.48 | - |
| +20% | 0 | 0 | 0.31 | 1.00 | 0 | 0.52 | |
| +40% | 0 | 0 | 0.53 | 1.00 | 0 | 0.55 | - |
| +60% | 0 | 0 | 0.75 | 1.00 | 0 | 0.58 | 38 |
| +80% | 0.02 | 0.14 | 0.80 | 1.00 | 0 | 0.91 | 33,34,36,38,40, 41,42 |
| +100% | 0.03 | 0.31 | 0.84 | 1.00 | 0.10 | 0.96 | 33,34,35,36, 37,38,39,40,42 |

TABLE VII
MICROGRID COSTS WITH RESPECT TO LOAD CHANGES

| Load change percentage | Investment Cost ($) | Operation Cost ($) | Planning Cost ($) |
|---|---|---|---|
| Orig. Load | 1310805 | 8151773 | 9462578 |
| +20% | 1497853 | 9961260 | 11459110 |
| +40% | 1683933 | 11734030 | 13417960 |
| +60% | 1876168 | 13594840 | 15471010 |
| +80% | 2644088 | 14884100 | 17528190 |
| +100% | 2904977 | 16729010 | 19633980 |

**Case 3:** In this case, market prices are changed from -80% to +80%, and their impact on planning results is studied. Price changes are studied in two cases of considering and ignoring line installations. The planning results are represented in Tables VIII and IX. By 20% decrease in the market price, the total dispatchable capacity would remain unchanged, but unit 1 is also installed as its capital cost is relatively low, and as the market price has decreased. Therefore, it is financially beneficial to dispatch unit 1 instead of increasing the installed capacity of units 3 and 4. By 40% decrease in market prices, units 3 and 4 are not installed anymore, but unit 1 is installed with a higher capacity, as its annualized investment cost is lower. By additional reduction in the market price up to -80%, the solar unit is not installed either, and only a capacity of 1.09 MW of unit 1 is installed. This is due to the very low market price which makes it more economical to buy power from the utility grid. Unit 1 is installed in this case for the mere purpose of supplying critical loads. On the other hand, by increasing the market price, it would be desirable for the microgrid to generate more power in order to sell back to the utility grid (associated with negative exchange power with the utility grid in many hours), which causes the operation cost to drop. Therefore, more DG capacity would be installed. For more than 40% increase in the market price, a number of distribution lines become congested, therefore, candidate lines are also installed, which would cause an increase in the investment cost. By increasing the market price, the decrease in the operation cost would be more dominant over the increase in the investment cost, so the planning cost would increase and then decrease. As shown in Table IX, for the values of percentage change in market prices that candidate lines are installed, the planning cost is lower than the case without allowing the installation of candidate lines. It means that the simultaneous installation of

DGs and candidate lines would be economically more viable than installation of only DGs. Also, it should be mentioned that the DES is not installed for any change in market prices.

TABLE VIII
INVESTMENT PLAN WITH RESPECT TO MARKET PRICE CHANGES

| Price change percentage | | 1 | 2 | 3 | 4 | 5 | 6 | Installed Lines |
|---|---|---|---|---|---|---|---|---|
| -80% | w/o lines | 1.09 | 0 | 0 | 0 | 0 | 0 | - |
|  | w/ lines | 1.09 | 0 | 0 | 0 | 0 | 0 | - |
| -60% | w/o lines | 1.09 | 0 | 0 | 0 | 0 | 0 | - |
|  | w/ lines | 1.09 | 0 | 0 | 0 | 0 | 0 | - |
| -40% | w/o lines | 1.09 | 0 | 0 | 0 | 0 | 0.48 | - |
|  | w/ lines | 1.09 | 0 | 0 | 0 | 0 | 0.48 | - |
| -20% | w/o lines | 0.45 | 0 | 0 | 0.64 | 0 | 0.48 | - |
|  | w/ lines | 0.45 | 0 | 0 | 0.64 | 0 | 0.48 | - |
| Orig. Price | w/o lines | 0 | 0 | 0.10 | 1.00 | 0 | 0.48 | - |
|  | w/ lines | 0 | 0 | 0.10 | 1.00 | 0 | 0.48 | - |
| +20% | w/o lines | 0.32 | 0 | 0.67 | 1.00 | 0.12 | 0.48 | - |
|  | w/ lines | 0.32 | 0 | 0.67 | 1.00 | 0.12 | 0.48 | - |
| +40% | w/o lines | 0.46 | 0 | 0.68 | 1.00 | 0.12 | 0.48 | - |
|  | w/ lines | 0.75 | 0 | 0.68 | 1.00 | 0.21 | 0.77 | 34,35,36,39,41 |
| +60% | w/o lines | 0.50 | 0 | 0.68 | 1.0 | 0.12 | 0.47 | - |
|  | w/ lines | 0.79 | 0 | 0.68 | 1.0 | 0.21 | 0.77 | 34,35,39,41 |
| +80% | w/o lines | 0.62 | 0 | 0.68 | 1.0 | 0.19 | 0.44 | - |
|  | w/ lines | 0.88 | 0 | 0.68 | 1.0 | 0.30 | 0.73 | 34,35,39,43 |

TABLE IX
MICROGRID COSTS WITH RESPECT TO MARKET PRICE CHANGES

| Price change percentage | | Investment Cost ($) | Operation Cost ($) | Planning Cost ($) |
|---|---|---|---|---|
| -80% | w/o lines | 510117 | 1957936 | 2468052 |
|  | w/ lines | | | |
| -60% | w/o lines | 510117 | 3915872 | 4425988 |
|  | w/ lines | | | |
| -40% | w/o lines | 1106618 | 5270609 | 6377227 |
|  | w/ lines | | | |
| -20% | w/o lines | 1225734 | 6883798 | 8109532 |
|  | w/ lines | | | |
| Orig. | w/o lines | 1310805 | 8151773 | 9462578 |
|  | w/ lines | | | |
| +20% | w/o lines | 1988144 | 8298427 | 10286570 |
|  | w/ lines | | | |
| +40% | w/o lines | 2059625 | 8622853 | 10682480 |
|  | w/ lines | 2824677 | 7354827 | 10179500 |
| +60% | w/o lines | 2078623 | 8893008 | 10971630 |
|  | w/ lines | 2831566 | 7307358 | 10138920 |
| +80% | w/o lines | 2174184 | 9012843 | 11187030 |
|  | w/ lines | 2883257 | 7110567 | 9993824 |

## IV. CONCLUSION

A microgrid co-optimization generation and distribution planning was proposed in this paper, with the objective of determining the optimal DER generation mix and upgrading the network by building new lines. The nonlinear power flow equations were linearized to formulate the problem by MILP. The problem was tested on the IEEE 33-bus standard system, demonstrating the sensitivity of the planning results with respect to various planning factors, including the ratio of critical loads, total aggregated load, and electricity prices. Obtained results advocated that microgrid planners can ensure better planning economics by considering a simultaneous expansion in generation and distribution as opposed to traditional models focused only on generation expansion.